\documentstyle{article}

\begin{document}

\title{MONTE CARLO SIMULATION OF SOME DYNAMICAL ASPECTS OF DROP
  FORMATION}

\author{A. R. de Lima.\thanks{E-mail: arlima@if.uff.br} , T. J. P.
  Penna\thanks{E-mail: tjpp@if.uff.br}\,\,
  and P. M. C. de Oliveira\thanks{-E-mail: pmco@if.uff.br}\\
  {\it Instituto de F\'\i sica - Universidade Federal Fluminense}\\
  {\it Av. Litor\^anea, s/n.$\!\!^\circ$ - Gragoat\'a}\\
  {\it Niter\'oi, Rio de Janeiro 24210-340, Brazil }}

\maketitle

\begin{abstract}
  In this work we present some results from computer simulations of
  dynamical aspects of drop formation in a leaky faucet. Our results,
  which agree very well with the experiments, suggest that only a few
  elements, at the microscopic level, would be necessary to describe
  the most important features of this system. We were able to set all
  parameters of the model in terms of real ones. This is an additional
  advantage with respect to previous theoretical works.
\end{abstract}

\vspace{0.5cm}

 
\section{Introduction}
        
Many experiments [1-6] have confirmed that the dripping faucet, in
spite of being a very simple system, presents complex behavior (chaos,
Hopf bifurcations, long range anti-correlations, hysteresis,
intermittence,...). Some years ago, a phenomenological macroscopic
model was introduced \cite{shaw84} to explain the dynamical properties
of this system based in the time interval between two drops. It was
proposed that the drop oscillates like a mass-spring system, and the
equation which represents it is given by
\begin{equation}
\label{eqshaw}
\frac{d}{dt}\left(m\frac{dy}{dt}\right)=mg-ky-b\frac{dy}{dt}
\end{equation}
\noindent
where $m$ is the mass of the forming drop, $y$ is its position and
$g$, $k$ and $b$ are constant parameters. This model reproduces some
return maps similar to experimental ones, but it cannot give any
information about, for instance, the drop shape and, in addition, the
parameters could not easily be compared to the experimental ones. From
a stochastic 2D boolean microscopic model in a square lattice,
originally applied to determine the shape of a drop on the wall
\cite{manna92}, some dynamical aspects of a growing drop were
reproduced recently\cite{deoliveira93,deoliveira94}. These findings
suggest that it is possible to obtain informations about the time
series and drop shapes as a dynamical process. Besides this, other
applications of lattice models to simulate macroscopic flow have been
shown to be very useful. Some examples are the FHP cellular automata,
which approximate the Navier-Stokes equations \cite{frisch86}, and the
study of mechanisms of fluid spreading with Ising model simulations
\cite{lukkarinen95}.

In a recent experimental paper\cite{sartorelli96}, two dimensional
images of growing drops were obtained and studied by image processing
techniques. In this work we show that from these studies, we succeeded
in relating the computational parameters of the stochastic model
\cite{deoliveira93,deoliveira94} to the real ones. We also discuss
some morphological features of the drop formation similar to the ones
exhibited in laboratory.

The main advantage of this lattice model in comparison with the
differential-equation approach is that we can obtain information about
the shape and, also, about time series with a large number of drops
without huge efforts. However, in this work, we did not perform any
smoothing of the drop surface because we are just interested in
studying a model which displays the same dynamical behavior of real
drops. It could be easily done if we are interested in the static
shape of drops in a dripping faucet.
        
This paper is organized as follows: in sec. 2 we explain some features
of the original model \cite{deoliveira93,deoliveira94} and the
modifications needed to simulate wider faucets. In sec. 3 we present
the results and conclusions will be discussed in sec. 4.
        
\section{The Model}
\label{themodel}
\noindent 
As in the original model \cite{manna92,deoliveira93}, we deal with a
square lattice where the state of each site can be represented by
Ising variables $\sigma=\pm 1$, where $+1$ means the fluid and $-1$
the air. In order to define the dynamics we introduced a hamiltonian
function
\begin{equation}
\label{hamiltonian}
{\cal H}=-\sum_{nn}\sigma_{i}\sigma_{j}-\sum_{nnn}\sigma_{i}\sigma_{j}-\sum_{j}h_{j}\sum_{row\,j}\sigma_{i}
\end{equation}
\noindent
where the first summation is over the nearest neighbors, and the
second one is over the next-nearest neighbors. These terms play the
role of both the molecular attraction and surface tension effects. The
third summation refers to the presence of an external potential
(gravity in the real system) varying linearly with the height, i.e,
$h_j=g \cdot j$, where $g$ is a constant and $j=1...V$, $V$ is the
vertical lattice size. A spin-flip Monte Carlo dynamics is imposed and
it is performed by Kawasaki trials (pair updates) imposing mass
conservation. This is made chosing at random one element (air) in the
external boundary and another one in the inner boundary of the drop
(fluid). The flip of both elements do not change the mass but it can
cause a variation of the energy as defined in equation
(\ref{hamiltonian}). Hereafter, we call ``relaxation'' each pair
update.
 
In the original implementation \cite{deoliveira93}, there are $W$
elements of fluid in the center of the first row of the lattice,
representing the width of the faucet. There, the injection of fluid
takes place, from the faucet into the drop, moving the drop one row
down without changing its current shape. Here we choose a different
approach. We inject $F$ elements on the outer boundary, and relax the
drop $N \times F$ times; i.e, $N \times F$ pair updates are made. In
this way the drop movement will be governed only by the Hamiltonian
$\cal H$. Time is defined proportional to the number of injections,
and it is independent of the drop perimeter, other difference
concerning the original model. This will allow us to compare drops at
different regimes of flux. Then we can keep the ``faucet width'' ($W$)
as constant changing only the flux ($F$) of fluid into the system to
compare with the experimental data.

A fundamental modification refers to the way we can determine when the
drop is disconnected from the top. Previously a burning algorithm was
used, spending much computer time. Since we are working with large
drops the disconnection is determined simply by searching for the
existence of an empty row (at which all spins are zero) on the
lattice, followed by a non empty row (that has at least one non-zero
spin) at each drop relaxation. By a visual inspection we observe that
inclined disconnections was not obtained (we consider only the zero
temperature).

As we intend to show below, these modifications were important in
order to relate the artificial parameters to the real ones.

\section{Results}
\label{results}
\noindent

Our typical system size is a $128\times460$ square lattice (our 32 bit
computer stores a $4\times460$ array, using $7360$ bytes). This
lattice size is approximately 12 times bigger than previous ones. We
carried out our simulations on 486DX4 100Mhz microcomputers with 8 MB
of memory. The code was compiled with GNU C using LIBGRX for
graphics\cite{code}. The CPU time is strongly dependent of the flux
(as in the experiments, the smaller the flux, the larger the time
required for a drop to disconnect from the faucet). In a typical run a
time series of $500$ drops ($W=40$, $F=5$) can be obtained after one
hour.

We consider sets with $500$ drops and $g=0.1$, $N=150$ and $W=40$. The
$100$ initial drops in each set were discarded to avoid transient
effects. All our results will be compared with the experimental
ones\cite{sartorelli96}, where two-dimensional images of growing drops
are reported.

In Fig. (1) we show some steps of the drop formation for some values
of $F$ (flux) to be compared with the real ones \cite{sartorelli96}.
In the experiment the increasing volume of the drop is found to be
proportional to time. The simulation presents the same qualitative
behaviour, as shown in Fig. (2). Just like in ref 12, it was observed
that the time evolution of the mean volume is

\begin{equation}
\label{volume}
<V>\,=V_{0}\,+<\phi>t
\end{equation}

\noindent
where $<\phi>$ is the mean water flux, equivalent to our $F$. From
this result we can verify that our redefinition of time is precise,
whereas considering the number of relaxations to be proportional to
the perimeter of the drop\cite{deoliveira94} do not reproduce this
behavior.

We can verify the non-linear character of the drop formation looking
at the center of mass height ($y$) as a function of time. This is
presented in Fig. (3) for three successive drops, for two different
values of $F$. As suggested in the experimental work, the drop
presents a sort of ``elastic behaviour'' evident at beginning, while
the center of mass height is still proportional to the time. After
that the drop shows a ``plastic deformation'', with a drastic change
in the center of mass height. The current analogy is made in terms of
the stress of some material under constant pressure, where after a
well defined limit the material is deformed in a ``plastic regime''.
Another interesting aspect is that, as in the laboratory, the drops at
higher flux regimes break down in a smaller time interval. This
behaviour is also evident in Fig. 4, where we have the center of mass
height ($y$) $\times$ drop volume, to all drops on each set.

On Figs. 5 and 6, we choose one drop and show its area as a function
of perimeter and perimeter over area as a function of time. These
plots show that the simulated drops present the same relations between
these variables which characterize the shape of the real drops. Once
more the behaviour are qualitatively identical to the experimentally
observed \cite{sartorelli96}. It is worth to note the time at neck
formation, where also these curves change drastically their behaviors.

In summary, we have show that $F$ is closely related to the flux and
$W$ is a fixed parameter which represents the faucet width. The
parameter $N$ (number of relaxations) is associated to the way the
fluid relaxes or, in physical terms, its viscosity. Small values of
$N$ are associated with small viscosities. Now we will comment about
the role played by the parameter $g$. In the previous
work\cite{shaw84,deoliveira93,deoliveira94}, different attractors have
been reached with changes in $g$. But a simple dimensional analysis of
${\cal H}$ shows that the last sum is the gravitational potential
energy. If we imagine that each Ising variable $\sigma_i$ represents a
small element of volume $dV$, thus $g$ can only be associated to the
weigth of that volume element. Since we have {\it
  weigth=density$\times$gravity$\times$dV}, $g$ can be associated to
the fluid density or even the gravity.

\section{Conclusions}
\label{conclusions}
\noindent

We showed that a 2D Ising-like model can simulate the drop formation
in a leaky faucet, both in the dynamical aspects and in the
morphological ones as well. We relate all the simulation parameters
with the real ones: $F$ with flux, $N$ with drop relaxation
(viscosity), $g$ with the weight of a volume element represented in
the model by the Ising variables and $W$ is associated to the width of
the real faucets.

We could show that only a few elements, within a microscopic approach,
may be enough to formulate a good theoretical model. Vibrations and
instabilities in the drop surface do not seem important to complex
behaviour of drops, from dripping faucets, since these ingredientes
are not taken into account in the present model. In a computational
point of view, lattice methods are very fast and they can give us many
additional informations as the ones from long time series, for
example, which are very difficult to be obtained by the traditional
theoretical methods, and of difficult control in the experiments.

It was presented in a recent paper\cite{penna95} that time series from
heartbeats present similar behaviour to that of leaky faucets.
Therefore, we suggest that this lattice model could be used to provide
more informations about the dynamics of the heart. Another immediate
application is to generate long temporal series and to study the
capacity of the chaotic prevision method applied to this complex
regime.  Works along these lines are in progress.

\section{Acknowledgments}
\noindent

This work was partially supported by Brazilian agencies CNPq, FINEP,
FAPERJ and CAPES. The authors thank H. J. Herrmann for discussions.

\newpage
\section*{Figure Captions}
\noindent
Fig. 1 - Some steps of one drop formation. The numbers at left are the
values of $F$ (flux) used in the simulation. We did not perform any
average to smooth their boundaries and the interval between the frames
are not the same.

\vspace{1cm}
\noindent 
Fig. 2 - Volume as function of time for the last drop in the sets with
$F=5$ ($\diamond$), $F=20$ ($\circ$) and $F=30$ ($\bullet$). The
linear behavior shows that our definition is in agreement with the
real one.

\vspace{1cm}
\noindent
Fig. 3 - Center of mass height of three successive drops, as function
of time, for two different values of $F$ ($F=20$($\circ$) and
$F=30$($\bullet$)). We can see that in regime with a lower $F$ the
drops spend more time to break down, exactly as in the experiments.

\vspace{1cm}
\noindent
Fig. 4 - Volume as function of the center of mass height to all drops
in the sets with $F=5$(a), $F=10$(b) and $F=20$(c). Again, the
simulation agrees with the experimental results, since an initially
linear behaviour was found. The saturation region (corresponding to
neck formation) is also observed.

\vspace{1cm}
\noindent 
Fig. 5 - Drop area vs perimeter. We present the results for only one
drop at the set $F=5$. Similar plots are found for all other drops.

\vspace{1cm}
\noindent
Fig. 6 - Ratio drop perimeter/area as a function of time. We consider
the same drop as in Fig. 5 for the $F=5$ set.

\end{document}